\documentclass[draft]{agujournal2019}
\usepackage{url}
\usepackage{amsmath}
\usepackage{soul}

\draftfalse

\journalname{JGR: Planets}

\begin{document}

\title{Dust Devils on Titan}

\author[1,*]{Brian Jackson}
\author[2]{Ralph D.~Lorenz}
\author[3]{Jason W.~Barnes}
\author[1]{Michelle Szurgot}
\affiliation{1}{Department of Physics, Boise State University, 1910 University Drive, Boise ID 83725-1570 USA}
\affiliation{2}{Applied Physics Laboratory, Johns Hopkins University, Laurel MD 20723-6099 USA}
\affiliation{3}{Department of Physics, University of Idaho, Moscow ID 83844-0903 USA}

\correspondingauthor{Brian Jackson}{bjackson@boisestate.edu}

\begin{keypoints}
\item As probed by Hyugens, the meteorological conditions near the surface of Saturn's moon Titan appear conducive to dust devil formation.
\item Dust devils may contribute significantly to Titan's aeolian cycle.
\item If dust devils are active on Titan, NASA's upcoming Dragonfly mission is likely to encounter them.
\end{keypoints}

\begin{abstract}
Conditions on Saturn's moon Titan suggest dust devils, which are convective, dust-laden plumes, may be active. Although the exact nature of dust on Titan is unclear, previous observations confirm an active aeolian cycle, and dust devils may play an important role in Titan's aeolian cycle, possibly contributing to regional transport of dust and even production of sand grains. The Dragonfly mission to Titan will document dust devil and convective vortex activity and thereby provide a new window into these features, and our analysis shows that associated winds are likely to be modest and pose no hazard to the mission.
\end{abstract}

\section*{Plain Language Summary}
Saturn's moon Titan may host active dust devils, which are small dust-laden plumes, which could significantly contribute to transport of dust in that moon's atmosphere. Although the exact nature of dust on Titan is unclear, previous observations confirm there is actively blowing dust on that world. If dust devils are active on Titan's surface, NASA's upcoming Dragonfly mission is likely to encounter them, but dust devils on Titan are unlikely to pose a hazard to the mission.

\section{Introduction}
Saturn's moon Titan is the only satellite in our solar system with a significant atmosphere -- consisting primarily of nitrogen, the atmosphere has a surface pressure 50\% larger than Earth's \cite{2005Natur.438..785F}. Titan also exhibits active hydrology, resulting in Earth-like river channels, gorges, and even lakes \cite{2018NatGe..11..306H}, and the Cassini Mission orbiting Saturn from 2004 to 2017 observed large, vigorous convective storms \cite{2011Sci...331.1414T}. At Titan's temperatures, water forms the bedrock, while atmospheric methane and ethane act as condensables. Titan's atmosphere is also laden with organic aerosols, produced high into the thermosphere by photolysis of N$_{2}$ and CH$_4$ and recombination into long hydrocarbon chains. These aerosols slowly settle out onto Titan's surface at a rate of about $10^{-7}\,{\rm kg\ m^{-2}\ s^{-1}}$   \cite{2014Icar..243..400L}.

Titan also exhibits active aeolian processes. Cassini's instruments observed fields of giant sand dunes girding Titan's equator and covering about 13\% of its surface \cite{2008Icar..194..690R}. How the dune sands, with expected grain diameters $D_{\rm p} \sim 200\,{\rm \mu m}$ \cite{2015PlSci...4....1B}, and dust grains, with diameters $D_{\rm p} \sim 5\,{\rm \mu m}$ \cite{2018NatGe..11..727R}, are produced is unclear, but their organic composition probably ultimately derives from the atmospheric aerosols \cite{2015PlSci...4....1B}. Cassini's Visual and Infrared Mapping Spectrometer (VIMS) also revealed three regional dust storms near the equator \cite{2018NatGe..11..727R} that occurred during the equinox season in 2009. Radiative transfer modeling implicate dust grains about $5\,{\rm \mu m}$ in radius and resembling the organic haze in composition \cite{2018NatGe..11..727R}. General Circulation Models (GCMs) of Titan predict precipitation at low latitudes during this season as the result of seasonal convergence of Hadley cells \cite{2019Icar..333..113L}. The strong winds associated with this seasonal precipitation likely drive the dust storms at these low latitudes where the dune fields reside.

The dust storms' appearance not only confirms an active dust cycle, but they also suggest the possibility of other dust-lifting phenomena. Among the most ephemeral but frequent such processes on other worlds are convective, dust-laden vortices, called dust devils. Dust devils occur throughout arid regions on Earth, where they contribute to poor air quality \cite{1990AtmEn..24.1135G}, and ubiquitously on Mars, where they may pose a hazard for human exploration \cite{2006RvGeo..44.3003B}. While dust devils observations go back millenia \cite{2016SSRv..203....5L}, our understanding of their operation and how they lift dust remains poor. If dust devils do exist on Titan, the unique aerodynamic environment may provide an unparalleled window into dust devil physics. 

The recently selected Dragonfly Mission could serve to elucidate dust devil processes through extended observation on the ground (rather like Mars landers). Dragonfly is a rotorcraft lander that will fly on Titan to explore that world's potential habitability and study its complex methane cycle. Titan's thick atmosphere and low gravity facilitate flight, motivating proposals to explore Titan by air for two decades \cite{2000JBIS...53..218L, 2005P&SS...53..606S, 2012ExA....33...55B}. The Dragonfly Mission was selected for launch in 2026 and arrival at Titan by 2034 \cite{Lorenz2018} and will fly over Titan's equatorial sand seas and characterize the chemical, geophysical, meteorological, and aeolian environment \cite{2018EGUGA..2019456R}. As we argue here, dust devils may stalk these very dunes, and so Dragonfly is likely to encounter active dust devils, most while landed, allowing us to probe their structures.

For this study, we focus on meteorological conditions near Titan's equator, where sand dunes and dust storms have been observed. Although dust devils may occur elsewhere on Titan and, as we argue, may contribute to global dust transport, the lack of near-surface data for other latitudes precludes a more comprehensive analysis. In addition, many of the relationships used throughout this study apply to convective vortices in general, not just dust devils. Of particular importance, Equation \ref{eqn:fractional_area} used to assess dust devil frequency considers all convective plumes, not just dust devils. We argue that winds in some vortices are likely sufficient to loft dust, and we attempt to account for dustless vortices. However, such vortices very likely occur alongside dust devils on Titan, if either occurs, and the relative occurrence rates are unclear. This is a key uncertainty in our study and may be addressed by the Dragonfly Mission.

\section{Properties of Dust Devils on Titan}

\begin{figure}
    \centering
    \includegraphics[width=\textwidth]{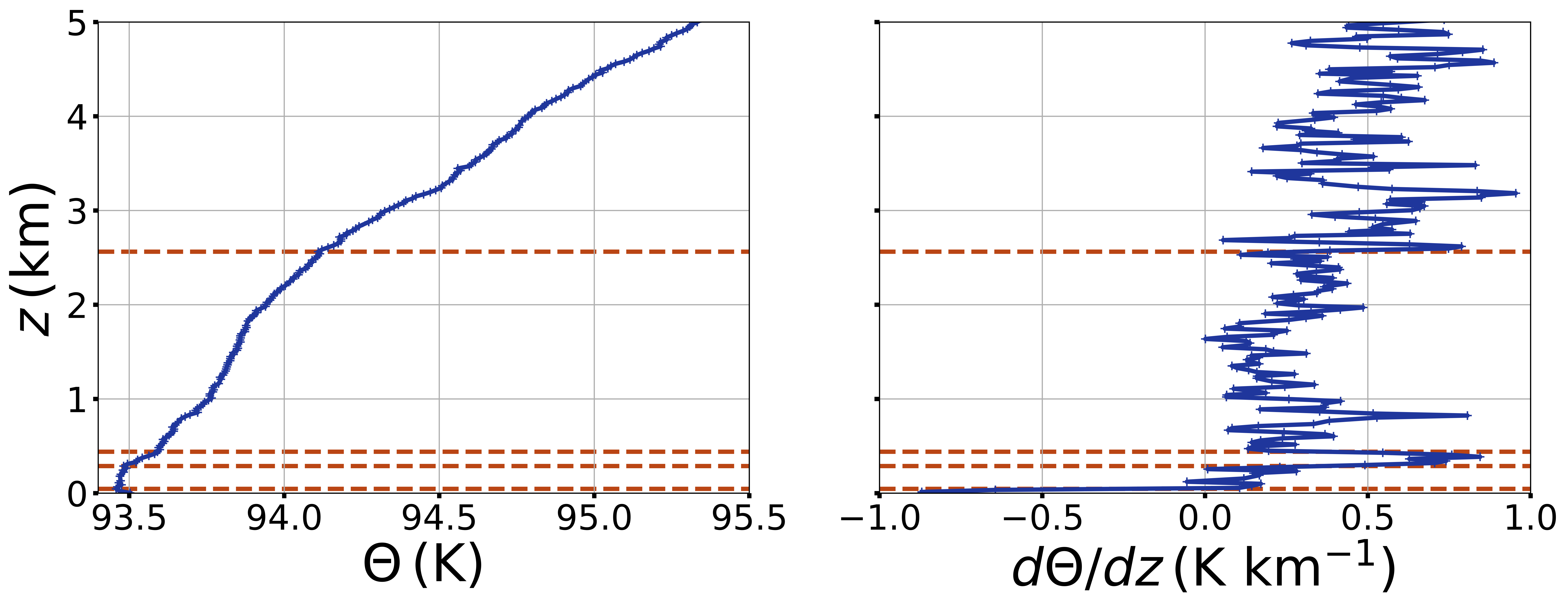}
    \caption{{\bf Titan's potential temperature profile and its derivative.} The left panel shows the inferred potential temperature profile $\Theta$ (blue) in degrees Kelvin as a function of altitude $z$ in km, while the right panel shows the derivative. The dashed orange lines indicate statistically significant kinks in the $\Theta$ profile, possibly corresponding to boundary layer features.} 
    \label{fig:PBL_analysis}
\end{figure}

Titan's meteorological conditions determine convective vortex and dust devil occurrence and properties (wind speeds, sizes, occurrence frequency, etc.). Several analytic and empirical criteria have been developed to relate these conditions to vortex and dust devil properties. The structure of the planetary boundary layer plays a key role, and it was probed at high resolution by the Huygens probe on 2005 January 14 at 09:47 local solar time \cite{2010cosp...38.1374C}. Figure \ref{fig:PBL_analysis} shows the potential temperature profile $\Theta$ calculated from Huygens measurements \cite{2006JGRE..111.8007T} of the actual temperature $T$ and atmospheric pressure $P(z)$ as $\Theta = T \left[P(z)/P_{\rm bot}\right]^\chi$, where $P_{\rm bot}$ is the atmospheric pressure at the bottom of the atmosphere (1470 hPa), and $\chi$ is the ratio of the specific gas constant to the specific heat capacity at fixed pressure, 0.3108 \cite{2006JGRE..111.8007T}. We explored this profile for boundary layer structures by calculating an altitude derivative and applying a first-order Savitzky-Golay filter \cite{1964AnaCh..36.1627S} and a window size of three points (i.e., we fit a piece-wise first-order polynomial, three points at a time) to moderately smooth the derivative. Otherwise, statistically insignificant excursions in the derivative would hamper our search for robust variations. We then applied the Bayesian blocks algorithm \cite{2013ApJ...764..167S} to the derivative to search for change points, which may correspond to boundary layer depth $h$. Frequent convective overturn takes place in the boundary layer, which can produce a region with a uniform potential temperature (i.e., an isentropic region). Thus, a change in the potential temperature profile may reflect the top of the planetary boundary layer, as suggested by previous studies \cite{2006JGRE..111.8007T}.

Below 5-km altitude, we found four distinct change points at 46 m, 288 m, 440 m, and 2.6 km altitude. The lowermost change point may be the top of the surface layer \cite{1988aitb.book.....S}, while the next two points higher up coincide closely with the top of a boundary layer previously identified at $300\,{\rm m}$ \cite{2006JGRE..111.8007T} and which may diurnally deepen to $800\, {\rm m}$ \cite{2012NatGe...5..106C}. The point at 2.6 km may correspond to the seasonally averaged boundary layer suggested in analyses of GCMs  \cite{2012NatGe...5..106C} and the morphology of Titan's equatorial dunes \cite{2010Icar..205..719L}. For our analysis, we consider  440 m and 2.6 km for $h$, although it is possible the change point at 288 m also represents an atmospheric structure. Additional modeling is probably required to assess that possibility.

\begin{figure}
    \centering
    \includegraphics[width=\textwidth]{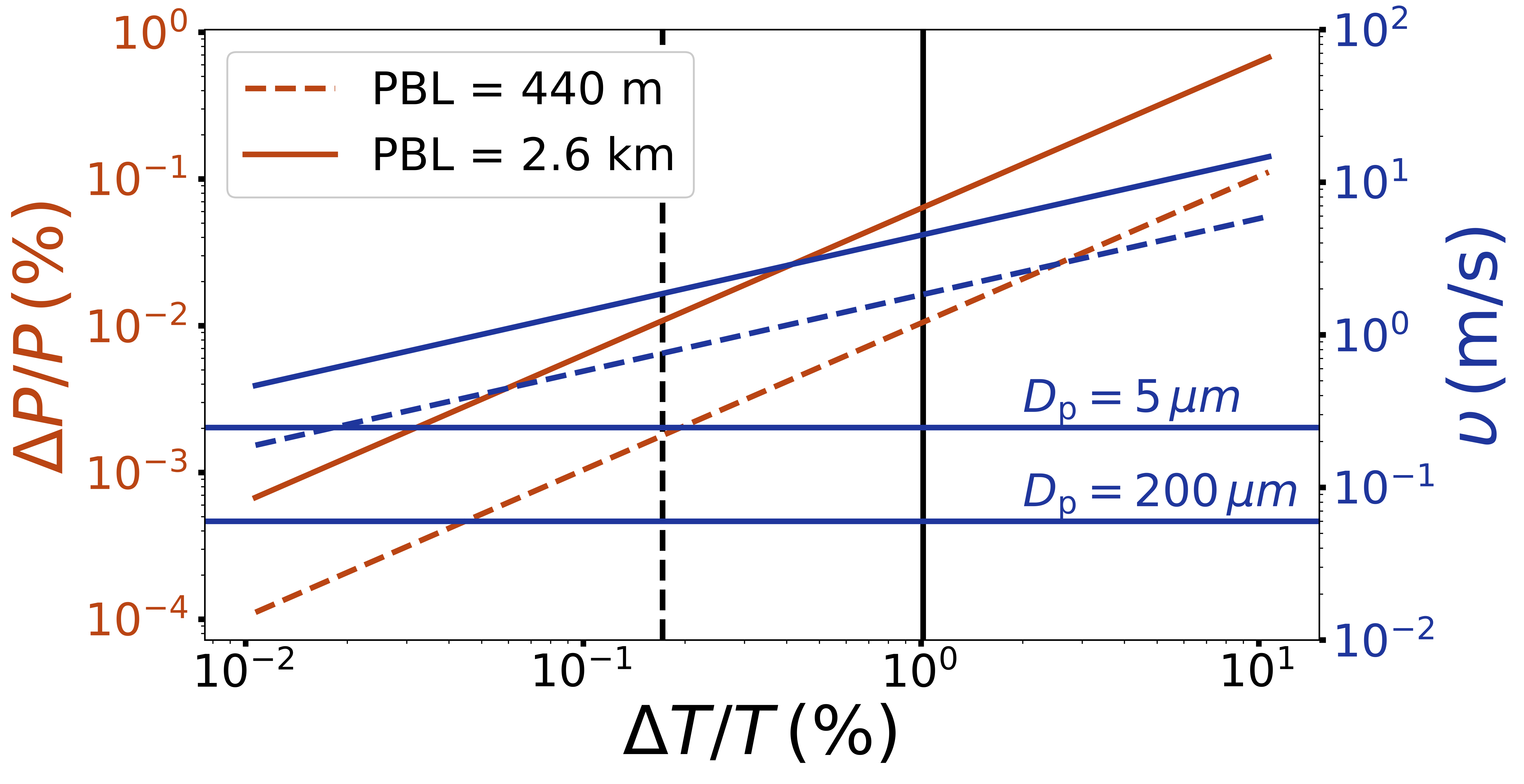}
    \caption{{\bf Dust devil pressure perturbations and wind speeds.} The fractional pressure perturbations $\Delta P/P$ (orange curves) and tangential wind velocities $\upsilon$ (blue curves) for a given temperature perturbation $\Delta T/T$ at the center of a convective vortex. The dashed lines assume a planetary boundary layer (PBL) of depth $440\, {\rm m}$, while the solid lines assume a depth of $2.6\,{\rm km}$. The solid, horizontal blue lines show the threshold wind speeds to initiate particle movement \cite{2015Natur.517...60B} with diameters $D_{\rm p} = 5\,\mu$m (dust) and $200\,\mu$m (sand). The black vertical lines show the minimum $\Delta T/T$ given by Inequality \ref{eqn:fractional_temperature_criterion}.}
    \label{fig:delta-P_and_tangential-wind}
\end{figure}

Dust devils can be modeled as Carnot heat engines \cite{1998JAtS...55.3244R} with an efficiency $\eta$ given by $\eta = \left( T_{\rm bot} - T_{\rm top} \right)/T_{\rm bot}$, where $T_{\rm bot}$ is the temperature at the surface (93.5 K as indicated by the Huygens' profile) and $T_{\rm top}$ is the temperature at the top of the boundary layer, which we take from Huygens' profile for a given value of $h$. The convective pressure perturbation at its center $\Delta P$ can be calculated as:
\begin{equation}
    \Delta P = P \left( 1 - \exp \left[ \left( \dfrac{\gamma\ \eta}{\gamma\ \eta - 1}\right) \left( \dfrac{\Delta T}{T_{\rm s}}/\chi \right) \right]\right),\label{eqn:Delta_P}
\end{equation}
where $\gamma$ is fraction of energy lost to friction with the surface ($\approx 1$) and $\Delta T$ is the entropy-weighted temperature contrast between the dust devil's convective center and the surrounding surface. Cyclostrophic balance in a convective vortex gives the tangential wind speed at the eyewall as $\upsilon^2 \approx \Delta P/\rho$, where $\rho$ is the atmospheric density ($5.3\,{\rm kg\ m^{-3}}$). Figure \ref{fig:delta-P_and_tangential-wind} shows the expected $\Delta P$ and $\upsilon$ for a range of $\Delta T$. The range of $\Delta T$ that actually manifests near Titan's surface is unclear, and additional modeling and direct observation are required for an accurate assessment. In absence of such information, for our calculations, we chose a range spanning well above and below previous estimates of $\sim 1\,{\rm K}$ \cite{2005Icar..173..222T}. The results show that for $h = 2.6\,{\rm km}$ and $\Delta T/T = 1\%$ ($\Delta T = 0.9\,{\rm K}$) that we expect $\Delta P/P \approx 6\%$ and $\upsilon \approx 5\,{\rm m s^{-1}}$, comparable to terrestrial dust devils \cite{2006RvGeo..44.3003B}.

In order to initiate a convective vortex, there must be sufficient convective energy in the boundary layer to overcome the boundary layer turbulence. Field work and theoretical analyses suggest the following criterion for dust devil formation \cite{2016SSRv..203..183R}:
\begin{equation}
    -\dfrac{h}{L} > 100,\label{eqn:Deardorff_criterion}
\end{equation}
where $h$ is the boundary layer depth and $L$ is the Obukhov length \cite{1971BoLMe...2....7O}. In this context, the Obukhov length is the height above the ground at which production of turbulence by shear and buoyancy are equal, and a small negative value may imply buoyant conditions \cite{2016SSRv..203..183R}. Based on the data returned by the Huygens lander \cite{2006JGRE..111.8007T}, $L$ was calculated as $-0.32\,{\rm m}$, giving $-h/L \approx 1000$ for $h = 440\,{\rm m}$. Inequality \ref{eqn:Deardorff_criterion} suggests a lower limit on $\Delta T/T$ required to form dust devils. Combining Inequality \ref{eqn:Deardorff_criterion} and the definition for $L$ (as described in detail in Appendix A), we find 
\begin{equation}
    \dfrac{\Delta T}{T} > \dfrac{100 \left( 4 f \right)^2}{\kappa g C_{\rm H}} h = \left( 3.9\times10^{-6}\,{\rm m^{-1}} \right)\ h,\label{eqn:fractional_temperature_criterion}
\end{equation}
where $f$ is the Coriolis parameter ($1.6\times10^{-6}\,{\rm s^{-1}}$), $\kappa$ the von K\'{a}rm\'{a}n parameter ($\approx 0.4$), $g$ the gravitational acceleration ($1.35\,{\rm m\ s^{-2}}$), and $C_{\rm H}$ the surface heat coefficient, which is comparable to the surface drag coefficient $\sim 0.002$ \cite{1977MWRv..105..215A, 2006JGRE..111.8007T}. The vertical black lines in Figure \ref{fig:delta-P_and_tangential-wind} show that the required variations are very modest for either value of $h$: less than $0.2\,{\rm K}$ for $h = 440\,{\rm m}$ and about $1\,{\rm K}$ for $h = 2.6\,{\rm km}$. Whether such temperature contrasts manifest over the scales relevant for convective vortex formation ($\sim 1\,{\rm km}$) is unclear since sufficiently high resolution temperature measurements are unavailable. However, models \cite{2005Icar..173..222T} of near-surface temperatures show that variations in surface albedo, emissivity, and/or thermal inertia can plausibly produce temperature variations $\sim 1\,{\rm K}$. As a check, we applied this criterion to terrestrial dust devils. For the Earth, $f \approx 8\times10^{-5}\,{\rm s^{-1}}$, $C_{\rm H}$ near the surface is $\sim 0.01$ \cite{1977MWRv..105..215A}, the mid-day boundary layer is typically $500\,{\rm m}$ deep \cite{Arya2001}, requiring $\Delta T/T > 0.1$. One field study of active terrestrial dust devils \cite{2003JGRE..108.5116T} found $\Delta T/T \sim 0.06$. 

We can explore convective instability by considering the Richardson number ${\rm Ri}$, defined as ${\rm Ri} = \left( g / T \right) \Gamma\ |\partial V/\partial z|^{-2}$, where $\Gamma$ is the atmospheric lapse rate, and $\partial V/\partial z$ is wind shear near the surface. A large, negative Richardson number ${\rm Ri}$ predicts convective instability \cite{Arya2001}. The near-surface potential temperature lapse rate measured during Huygens' descent was $\Gamma = -0.9\,{\rm K\ km^{-1}}$, and Doppler tracking of Huygens \cite{2005Natur.438..800B} found near-surface ($z < 660\,{\rm m}$) shear of about $10\,{\rm m/s/km}$ \cite{2017Icar..295..119L}. By fitting a first-order polynomial to the lowermost two points in the wind profile and considering the uncertainties, we estimate a shear of about $4\,{\rm m/s/km}$ and ${\rm Ri} = -0.8$ for the time and location probed by Huygens. Although the wind shear estimate here is based on only a few measurements, higher resolution wind profiles from GCM analyses \cite{2015Icar..250..516L} also predict a very small near-surface wind shear, which can help promote convective instability. Large eddy simulations of dust devil formation \cite{2016AeoRe..22...47K} suggest such a large and negative Richardson number corresponds to dust devil occurrence rates exceeding $1\,{\rm km^{-2}\ hr^{-1}}$ for Earth-like surface heating rates. Moreover, temperature and wind speed measurements made between 2 and 9-m altitude inside three active dust devils on the Earth \cite{1973JAtS...30.1599S} give ${\rm Ri} \le -0.4$.

Regarding the sizes of dust devils on Titan, imaging studies for the Earth and Mars suggest that dust devil diameters $D$ are about five times smaller than their heights \cite{2013Icar..226..964L} and their heights are capped by the PBL height \cite{2015Icar..260..246F}. Therefore, for $h = 2.6\,{\rm km}$, $D \sim 520\,{\rm m}$, and for $h = 440\,{\rm m}$, $D \sim 88\,{\rm m}$, small enough not to have been detected by Cassini. Regarding the smallest dust devils we might expect, the Obukhov length, $\sim 1\,{\rm m}$, may set the minimum diameter for convective vortices generally \cite{2011Icar..215..381L}. Our estimates for tangential velocities and diameters suggest vorticities $\zeta \approx 10^{-2}\,{\rm s^{-1}}$, much smaller than typical dust devils \cite{1998JAtS...55.3244R, 2011JGRD..11616120R} $\ge 1\,{\rm s^{-1}}$. However, since dust devils are known to span a wide range of sizes \cite{2016SSRv..203..277L}, this calculation may underestimate the typical vorticity. Indeed, previous studies \cite{2011Icar..215..381L} suggest terrestrial and martian dust devil diameters follow a power-law, $\delta(D) \propto D^\alpha$, with $-2 \le \alpha \le -1$ \cite{2016SSRv..203..277L}.

Dust devils do not involve latent heating and may, in fact, be inhibited by the presence of liquids. On Earth, wetting of the surface can suppress dust devils \cite{2015JGRE..120..401J} by increasing particle cohesion (and thus increasing the threshold wind required for dust-lifting) and surface thermal inertia, thereby damping the surface temperature response to solar heating and inhibiting boundary layer formation. Observations \cite{2005Sci...310..474G, 2011Sci...331.1414T, 2006Icar..182..224S, 2006Icar..184..517S} show that large methane downpours occur on Titan, but any given spot on the surface may only see rain once a ten Earth years \cite{2014Icar..241..269C}. Rainfall is predicted to re-evaporate over a period of a few Titan years \cite{2016Icar..267..106N}. Moreover, the presence of dunes \cite{2006Sci...312..724L} and dust storms \cite{2018NatGe..11..727R} near Titan's equator indicates insufficient surface wetting to totally extinguish aeolian processes. Surface humidity levels measured by Huygens were 50\%, too small for cloud formation, which may require a humidity level of 60\% \cite{2000Sci...290..509G}, and also too small for substantial rainfall, which may require a humidity level of 80\% \cite{2006Natur.442..428H}. Except for equinoctial (once every 16 years) outbursts at low latitudes \cite{2008JGRE..113.8015M, 2016Icar..267..106N, 2019Icar..333..113L}, Titan's global circulation drives condensables away from the equator \cite{2008JGRE..113.8015M, 2015Icar..250..516L}. Even though subsurface moisture was indicated by the Huygens instruments at its streambed landing site, the Huygens imager/spectrometer instrument detected possible dust lofted by the probe's aerodynamic wake at impact  \cite{2012P&SS...73..327S}. Thus, we expect arid conditions near the equator suitable for dust devil formation.

However, some conditions on Titan may impede dust devil production. The ambient wind field is probably the source for dust devil vorticity, requiring some minimum ambient wind speed; field studies \cite{2016SSRv..203..183R} suggest $1\,{\rm m\ s^{-1}}$. Whether the same threshold applies to Titan should be the subject of future work, but measured surface wind speeds \cite{2005Natur.438..800B} are $\le 1\,{\rm m\ s^{-1}}$, while results from \cite{2010AeoRe...2..113T} and \cite{2015Icar..250..516L} suggest typical speeds below $0.5\, {\rm m\ s^{-1}}$ at low latitudes. However, higher resolution models (and in particular Large Eddy Simulations, which have typical resolutions of order tens of meters) may produce different results, and therefore future work should consider such models. Moreover, additional work is required to assess whether a sufficiently large fraction of surface heating on Titan partitions into into sensible heating, which is required to drive convective vortex production, rather than latent heating. In other words, what is the relevant Bowen's ratio for Titan \cite{Arya2001}? Titan represents a key experiment in this aspect of micrometeorology. Static charging of grains may inhibit dust mobilization -- dust on Titan may develop much higher charges than terrestrial grains \cite{2017NatGe..10..260M}, forming $\sim{\rm mm}$-sized clumps which are much larger than expected for sand grains, and so mobilization may require wind speeds a few times larger than for uncharged grains. In the end, though, the presence of active dust storms on Titan means that surface winds must occasionally mobilize dust.

What might be the significance of dust devils for Titan's meteorology and aeolian cycle? On Mars, dust devils help maintain the background atmospheric haze, which probably contributes about $10\,{\rm K}$ of radiative heating \cite{2004JGRE..10911006B}. Continual high-altitude production of haze means Titan's atmosphere is already loaded with aerosols, and unlike on Mars, dust devils probably do not contribute significantly to maintaining the hazy atmosphere. However, dust devils may be important for aerosol transport since surface winds are usually weak. As on Mars, dust storms on Titan probably occur infrequently near the equator, perhaps during seasonal wind storms \cite{2018NatGe..11..727R}, but conditions may frequently allow dust devil formation. The results from lab experiments \cite{2006GeoRL..3319S09N} suggest that, for a PBL depth of $2.6\,{\rm km}$, a terrestrial dust devil with $\Delta T/T \sim 1\%$, and therefore $\Delta P/P \sim 0.1\%$, could lift $0.025\,{\rm kg\ m^{-2}\ s^{-1}}$. Given Titan's surface pressure, such a dust devil would probably lift more, greatly exceeding the haze production rate of $10^{-7}\,{\rm kg\ m^{-2}\ s^{-1}}$ \cite{2014Icar..243..400L}. Indeed, if they occur, dust devils may be the primary mode of dust transport in arid regions on Titan. They may also raise dust grains to altitudes where regional winds can pick them up and so contribute to regional dust transport, with ultimate deposition in Titan's seas \cite{2016SSRv..203..377K}. Dust devils may also figure in the creation of dune sands from dust grains. Tumbling in a dust devil may charge dust grains, and such charging can result in multi-particle clumps approaching $5\,{\rm mm}$ in size \cite{2017NatGe..10..260M}. Whether such clumps could survive reptation and saltation involved in dune evolution is unclear, but the clumping may help initiate the sand formation process \cite{2015PlSci...4....1B}. Interestingly, dust transport on Titan, where conditions may favor direct lifting of dust grains, may differ substantially from transport on Mars, where dust lifting is primarily via saltation since direct lifting on Mars requires very large wind speeds \cite{1985wagp.book.....G}.

\begin{figure}
    \centering
    \includegraphics[width=\textwidth]{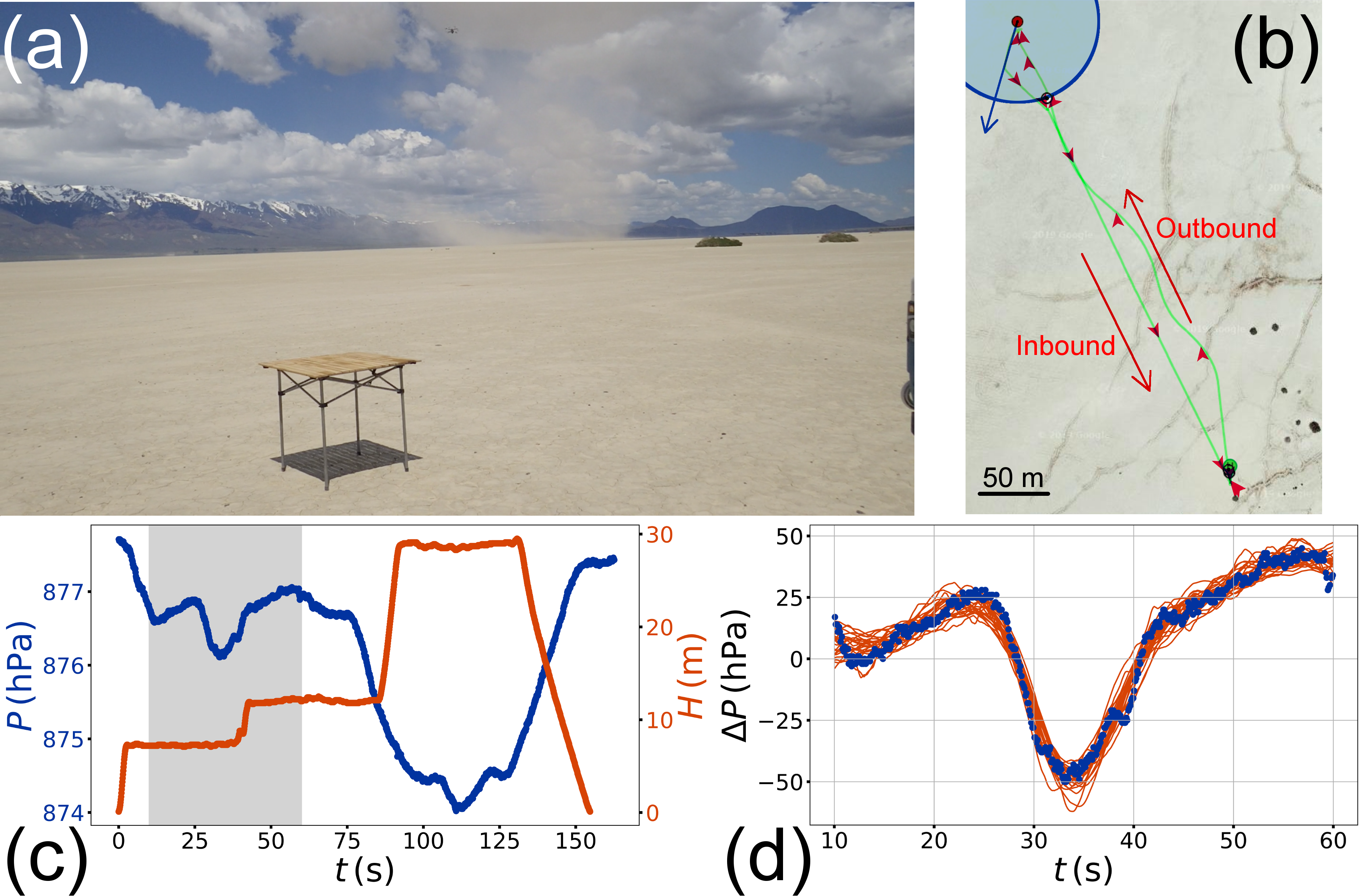}
    \caption{{\bf A dust devil encounter on the Alvord Desert in Oregon.} (a) Launch of the drone (visible at top middle) toward the dust devil. (b) Drone outbound and inbound trajectories, with dust devil's approximate size ($100\,{\rm m}$) and position at encounter depicted as a blue circle. (c) Pressure $P$ in hectoPascal shown in blue and drone altitude $H$ in meters shown in orange during the drone's flight. The grey shaded region highlights the dust devil encounter. (d) The pressure fluctuation associated with the dust devil in blue, along with model fits including a Gaussian process turbulent noise model \cite{2018RemS...10...65J} in orange. A video of the encounter is available at https://youtu.be/eOxklEkYwwE. \label{fig:May9_12-22p-Encounter}}
\end{figure}

For all its similarity to Earth, Titan represents a novel regime for aeolian studies. Given the wide availability of dust and the scaling arguments presented here, even the absence of dust devils would be an important result. Probing active dust devils with an instrumented drone has been the subject of recent successful terrestrial field studies \cite{2018RemS...10...65J} and is helping to reveal the relationships between pressure, temperature, wind, and a devil's dust-lifting capacity \cite{2016SSRv..203..347N}. Figure \ref{fig:May9_12-22p-Encounter} shows results from a recent field study conducted by our group. The pressure signal registered during the dust devil encounter is shown in panel (d). Fitting a standard Lorentzian profile \cite{2018RemS...10...65J}, we estimate $\Delta P = 80 \pm 11\,{\rm hPa}$, corresponding to a tangential velocity $\upsilon \approx \sqrt{80\,{\rm hPa}/1\,{\rm kg\ m^{-3}}} = 9\,{\rm m\ s^{-1}}$. By-hand feature-tracking analysis of the encounter video collected on-board the drone confirms $\upsilon \sim 10\,{\rm m\ s^{-1}}$. As we discuss in the next section, since Dragonfly will carry similar meteorological sensors and cameras, similar experiments can be conducted. However, since dust devils are most likely to be encountered while Dragonfly is on the ground, terrestrial \cite{2015JGRE..120..401J} and martian \cite{2006JGRE..11112S09G} fixed station studies of dust devils may be more analogous. For instance, collection of pressure and temperature time-series while being passed over by a dust devil will provide their profiles (although distinguishing dust devils from dustless vortices using only pressure and temperature is difficult). Analysis of dust devil image contrast can provide estimates of dust loading \cite{2006JGRE..11112S09G}. In cases where dusty and dustless vortices can be distinguished, post-processing of Dragonfly's pressure time-series may yield the signatures of dust devil encounters \cite{2015JGRE..120..401J} which, in turn, can provide the underlying occurrence rates \cite{2016SSRv..203..277L} and allow estimates of regional dust transport \cite{2016AeoRe..22...47K}.

\section{Discussion and Perspectives: The Dragonfly Mission}
\label{sec:Discussion and Perspectives}
Given the recent selection by NASA of the Dragonfly Mission to fly on Titan by 2034, we may have the opportunity to probe active dust devils in a totally novel environment. If dust devils exist on Titan, then the Dragonfly Mission can provide crucial insight into how they operate. The mission will probe the chemical, geophysical, meteorological, and aeolian environment \cite{2018EGUGA..2019456R}. The DraGMet geophysics and meteorology package will measure pressure, temperature, wind speed, and methane humidity, data which may elucidate structure and formation conditions for convective vortices including dust devils. DragonCam will provide panoramic imaging for navigation and may also provide images of dust devils and a microscopic imager to examine surface materials down to the scale of sand grains \cite{2019LPI....50.2885M}, crucial for assessing aeolian processes. Using $\sim 10\,{\rm km}$, $500\,{\rm m}$ altitude flights (at $\sim 10\,{\rm m\ s^{-1}}$ flight speed) powered by a Multi-Mission Radioisotope Thermoelectric Generator (MMRTG), Dragonfly will hop between equatorial interdunes to reconnoiter landing sites.

With wind speeds of a few$\,{\rm m\ s^{-1}}$, any convective vortex would only moderately perturb Dragonfly's flight, and, indeed, small, commercial quadcopters such as the DJI Phantom 4 (weighing $1.5\,{\rm kg}$ and spanning $50\,{\rm cm}$) are rated to fly in wind speeds up to $11\,{\rm m\ s^{-1}}$. A published estimate for Dragonfly's final mass exceeds 400 kg \cite{Lorenz2018}. As shown in Figure \ref{fig:delta-P_and_tangential-wind}, such wind speeds would require $\Delta T/T$ in excess of $10\%$ ($\approx 10\,{\rm K}$), much larger than the $1\,{\rm K}$ seen on Titan \cite{2005Icar..173..222T, 2012P&SS...60...62C, 2015Icar..250..516L}. More likely, for $h = 2.6\,{\rm km}$ and $\Delta T/T = 1\%$ ($1\,{\rm K}$), dust devils will exhibit $\upsilon \approx 4.5\,{\rm m\ s^{-1}}$. Since drag scales as $\rho \upsilon^2$ and Titan's atmospheric density is about five times Earth's, a $4.5\,{\rm m\ s^{-1}}$-wind on Titan is equivalent to a $10\,{\rm m\ s^{-1}}$-wind on Earth, below the threshold for much smaller commercial drones.

We can use our earlier estimates and adapt previous work to estimate the frequency of dust devil encounters for Dragonfly, both while the spacecraft is in flight and while it is in on the ground. The frequency of dust devil encounters will depend on the currently unknown flight trajectories and times of day and particularly on the dust devil areal density. We can estimate this density using scalings for convective activity that have previously been applied to Titan \cite{2005GeoRL..32.1201L}. The fractional surface area covered by convective plumes, $f$, (convective plumes of all types, not just dust devils) is given by
\begin{equation}
    f = \sqrt{\mu/\eta} \left( c_{\rm p} \Delta T \right)^{-3/2} \rho^{-1} F_{\rm in},\label{eqn:fractional_area}
\end{equation}
where $\mu$ is the ratio of frictional dissipation to a plume's kinetic energy and is $\approx 16$ \cite{1996JAtS...53..572R}. $F_{\rm in}$ is the surface heat flux. Plugging typical terrestrial values \cite{1998JAtS...55.3244R} into Equation \ref{eqn:fractional_area} gives $f_{\rm Earth} \approx 3\times10^{-3}$, but small-scale plumes such as dust devils comprise only one kind of convective system and therefore contribute only a portion of $f$. We can estimate the $f$-value relevant for terrestrial dust devils by considering power-law distribution of diameters (and therefore areas) and using results from in-situ visual surveys of dust devils \cite{2016SSRv..203..277L} (see Appendix A). With this approach, we infer a value for $f_{\rm Earth}$ about four times smaller than the value given by Equation \ref{eqn:fractional_area}. Turning to Titan, we can use the temperature profile in Figure \ref{fig:PBL_analysis} and, taking $h = 440\,{\rm m}$, $\eta = 0.003$. Into Equation \ref{eqn:fractional_area} we can plug $\Delta T = 1\,{\rm K}$ and $F_{\rm in} = 1\,{\rm W\ m^{-2}}$  \cite{2006JGRE..111.8007T} and scale down by the same factor of four as for the Earth, giving $f_{\rm Titan} = 10^{-4}$. We can then convert this fractional area coverage into an areal density for dust devils using
\begin{equation}
    N = \int_{D_{\rm min}}^{D_{\rm max}} f_{\rm Titan}\ \left( 4/\pi \right)\ D^{-2}\ \delta(D)\ dD,
\end{equation}
where $D_{\rm min/max}$ is the minimum/maximum diameter. With $D_{\rm max} = 440\,{\rm m}/5 = 88\,{\rm m}$ and $D_{\rm min} = L \approx 1\,{\rm m}$, $N \approx 28\,{\rm km^{-2}}$.

Assuming dust devils reach Dragonfly's altitude ($\approx 500\,{\rm m}$), an encounter while Dragonfly is in flight requires a devil to pass through a corridor centered on Dragonfly and running parallel to the flight path. The width of this corridor is equal to the devil's diameter; dust devils farther left or right will not register an encounter. We can estimate the mean free path between encounters for dust devils with a range of diameters as $\lambda = \int \left( N\ D \right)^{-1}\ \delta(D)\ dD = 13\,{\rm km}$ for $\delta(D) \propto D^{-1.5}$. This result is insensitive to the exact values for the power-law exponent and the minimum and maximum diameters. 

Assuming that Dragonfly's ground velocity during flight, $\sim 10\,{\rm m\ s^{-1}}$ \cite{Lorenz2018}, greatly exceeds the dust devils' velocities ($\sim 1\,{\rm m\ s^{-1}}$), this value for $\lambda$ would imply between two and three encounters during each flight. However, Dragonfly is planned to fly in the early morning, local Titan time, when conditions are calm and dust devils are probably less active. Since Dragonfly will spend most of its time on the ground, encountered dust devils will likely be carried by the ambient wind. Dust devil occurrence peaks when the surface temperature reaches a maximum \cite{2015JGRE..120..401J}, and observations of Titan's diurnal surface temperatures \cite{2012P&SS...60...62C} indicate a broad peak from about 11am till 3pm local time. Since Titan days are about 16 Earth days long, this timespan corresponds to 64 Earth hours. In that case, Dragonfly might encounter one dust devil every 3.6 Earth hours, or nearly 20 devils while the spacecraft is on the ground each Titan day. More detailed meteorological \cite{doi:10.1029/2019JE006082} and statistical \cite{2014JAtS...71.4461L} modeling of predicted dust devil activity as a function of time of sol and season could improve these estimates. However, in the absence of such modeling, these estimates suggest that, if dust devils occur on Titan, Dragonfly is likely to encounter them regularly.

\section{Conclusions}

As we detail here, meteorological conditions near the surface of Titan as probed by the Huygens lander and Cassini observations of dust storms suggest the possibility of active small-scale convective vortices including dust devils on that world. A boundary layer depth much larger than the Obhukov length scale \cite{2016SSRv..203..183R}, coupled to very low near-surface wind shear \cite{2017Icar..295..119L}, is consistent with active dust devils, even though temperature variations near Titan's surface (and therefore atmospheric buoyancy) are likely to be subdued compared to Earth and Mars \cite{2005Icar..173..222T}. 

If they exist, convective vortices including dust devils on Titan likely span a range of sizes, depending on the depth of the planetary boundary layer $h$ during the time of day they form, with typical diameters for dust devils of either $\sim 88\,{\rm m}$ (for $h = 440\,{\rm m}$) or $\sim 520\,{\rm m}$ (for $h = 2.6\,{\rm km}$; \citeA{2012NatGe...5..106C, 2010Icar..205..719L}). The tangential windspeeds at the vortex eyewalls are likely to be only a few ${\rm m\, s^{-1}}$, sufficient to lift the $5-{\rm \mu m}$ dust grains thought to populate Titan's dust storms \cite{2018NatGe..11..727R}. As winds near Titan's surface probably rarely exceed $1\,{\rm m\ s^{-1}}$ \cite{2005Icar..173..222T, 2012P&SS...60...62C, 2015Icar..250..516L}, dust devils may play a key role in Titan's aeolian cycle.

Field studies using instrumented drones have helped reveal interior structures and activity within terrestrial dust devils \cite{2018RemS...10...65J} and may provide key evidence to unravel the still-obscure physical mechanisms by which devils loft and carry dust. The arrival of NASA's Dragonfly Mission on Titan in 2034 presents the prospect of probing extraterrestrial vortices in a similar way to these field studies, although encounters will probably occur primarily while Dragonfly is on the ground. However, Titan's relevant meteorological conditions suggest that vortex encounters, if in-flight, may occur a few times during Dragonfly's daily flight. The low wind speeds expected for dust devils on Titan mean they will pose little to no hazard to the mission. However, Dragonfly will spend most of its time on the ground, including during Titan's mid-day when vortices are most likely to be active \cite{2015JGRE..120..401J}, and so encounters will probably occur on the ground every few Earth hours instead. In this case, they will likely resemble encounters on Mars by landed spacecraft. Even then, though, the imagery and meteorological data collected by Dragonfly during encounters may break new ground in aeolian studies by showing how they operate in a new aerodynamic environment.

\appendix
\section*{Appendix A}
\label{sec:Methods}
Here, we derive Inequality \ref{eqn:fractional_temperature_criterion}. It is important to note that Equations \ref{eqn:Delta_P}-\ref{eqn:fractional_temperature_criterion} in the main narrative above were originally developed specifically for application to dust devils, but they may also apply to convective vortices generally. In this Appendix, none of the equations apply uniquely to dust devils, although some of the numerical values (e.g., $R = 100\,\rm{km^{-2}\ day^{-1}}$) do.

To begin, Inequality \ref{eqn:fractional_temperature_criterion} relies on the definition of the Obukhov length, given by \cite{Arya2001} 
\begin{equation}
    L = \dfrac{-u_\star^3}{\left( \kappa g H \right)/\left( \bar{T} \rho c_{\rm p} \right) },\label{eqn:Obukhov_length}
\end{equation}
where $u_\star$ is the friction velocity, $\kappa$ the von K\'{a}rm\'{a}n parameter ($\approx 0.4$), $g$ the gravitational acceleration ($1.35\,{\rm m\ s^{-2}}$), $H$ the surface heat flux, $\bar{T}$ the average atmospheric temperature near the surface ($93.5\,$K), $\rho$ the atmospheric density at the surface ($5.3\,{\rm kg\ m^{-3}}$), and $c_{\rm p}$ the specific heat capacity ($1100\,{\rm J\ kg^{-1}\ K^{-1}}$). The friction velocity is related to the boundary layer depth via the Ekman layer approximation \cite{1995JGR...10026377L}:
\begin{equation}
    u_\star \approx 4 f h,\label{eqn:friction_velocity}
\end{equation}
where $f$ is the Coriolis parameter \cite{1992aitd.book.....H}, which we take as $1.6\times10^{-6}\,{\rm s^{-1}}$ since Dragonfly will land near Titan's equator. We can estimate the surface heat flux as \cite{Arya2001}
\begin{equation}
    H = C_{\rm H} \rho u_\star c_{\rm p} \Delta T_{\rm surf-air},\label{eqn:heat_flux}
\end{equation}
where $C_{\rm H}$ is the surface heat coefficient \cite{1977MWRv..105..215A} and $\Delta T_{\rm surf-air}$ is the difference in temperature between the ground and the near-surface atmosphere. Since the lateral temperature difference in the near-surface atmosphere $\Delta T$ that appears in Equation \ref{eqn:Delta_P} arises from the surface heating, we take $\Delta T \approx \Delta T_{\rm surf-air}$. Combining all these equations, plugging them into Inequality \ref{eqn:Deardorff_criterion}, and re-arranging gives Inequality \ref{eqn:fractional_temperature_criterion}.

Next, we discuss our calculation of the fractional area for terrestrial dust devils $f_{\rm Earth}$. Based on several in-situ visual surveys of dust devils, \cite{2016SSRv..203..277L} suggests that the dust devil occurrence rate in arid regions as integrated over all diameters is $R = 100\,{\rm km^{-2}\ day^{-1}}$. Therefore, the diameter power-law $\delta(D)$ must satisfy 
\begin{equation}
R = 100\,{\rm km^{-2}\ day^{-1}} = \int_{D_{\rm min}}^{D_{\rm max}}\ \delta_0\ D^{-1.5}\ dD,\label{eqn:N_integral}
\end{equation}
where $\delta_0$ is the normalization constant. Taking $D_{\rm min} = 1\, {\rm m}$ and $D_{\rm max} = 100\, {\rm m}$ \cite{2016SSRv..203..277L} and solving this equation gives $\delta_0 \approx 0.6\,{\rm km^{-2}\ day^{-1}\ km^{0.5}}$. The fractional area $f$ must satisfy
\begin{equation}
f = \int_{D_{\rm min}}^{D_{\rm max}} \delta\ A\ \tau\ dD,\label{eqn:fractional_area_definition}
\end{equation}
where $A$ is the area of a single dust devil, $\frac{\pi}{4} D^2$. The lifetime of a dust devil $\tau$ seems to scale with diameter \cite{2013Icar..226..964L} as $\tau = \left( 40\, {\rm s} \right) \left( D/{\rm m} \right)^{2/3}$. Plugging these expressions into Equation \ref{eqn:fractional_area_definition} and integrating gives $f_{\rm Earth} \approx 8.7\times10^{-4}$. It is important to note that visual surveys likely under-count small dust devils, those with little dust opacity, and completely miss dustless vortices.

\acknowledgments

The authors acknowledge considerable helpful input from the editor and our referees, Prof.~Jim Murphy and Dr.~Claire Newman. This work was supported by grant number 80NSSC19K0542 from NASA's Solar System Workings program and by a grant from the Idaho Space Grant consortium. RL acknowledges the support of NASA grants 80NSSC18K1626 (InSight PSP) and 80NSSC18K1389 (Cassini/Huygens). All data and the code to perform the calculations presented here are available at https://zenodo.org/record/3470280.

\bibliography{agusample}

\end{document}